\documentclass[12pt]{article}
\usepackage{amsmath}
\usepackage{amssymb}

\newcommand{\bb}{\begin{thebibliography}{99}}
\newcommand{\eb}{\end{thebibliography}}

\title{ SINGULAR STRUCTURE OF THE QED EFFECTIVE ACTION }
\author{ B.A.FAYZULLAEV\\
 Department of  Theoretical Physics,\\
 National University of Uzbekistan,\\
 Tashkent 100095, Uzbekistan}

\begin{document}

\begin{titlepage}

 \maketitle

\begin{abstract}
The equations for the QED effective action derived in \cite{fm} are
considered using singular perturbation theory. The effective action is divided into
regular and singular (in coupling constant) parts. It is shown that expression for the regular part
coincides with usual Feynman perturbation series over coupling constant, while the remainder
has essential singularity at the vanishing coupling constant: $e\rightarrow 0$. This means that in the frame of quantum field theory
it is impossible "to switch off" electromagnetic interaction in general and pass on to "free electron".

\end{abstract}
\vspace{.5cm} Key words: quantum electrodynamics, effective action,
singular perturbation expansion, vacuum expectations.
\thispagestyle{empty}
\end{titlepage}

\section{Introduction}
Investigation of analytic structure of quantum field models usually is reduced to  investigation of $S$-matrix as analytic function of its variables such as
angular momenta, energy,   masses of particles  etc.
Many results were reached in this way - see, e.g. \cite{pol}.
But there are no results concerning analyticity of quantum quantities, e.g.,  vacuum expectations or effective action.
Especially, analyticity  of vacuum expectations of quantum fields in coupling constants. Expanding amplitude of any process in perturbation  series over
coupling constant we implicitly suppose analyticity of this amplitude in this coupling constant. But it is well known that perturbation series in quantum field models
are asymptotic ones. In this article we want to investigate perturbation series in QED from the following viewpoint: is there any singularity in this series at $e=0,$ where $e$ is
electric charge (coupling constant). I.e., can we "switch off" the electromagnetic interaction in Quantum Electrodynamics?
In this article we show that this is impossible.

\section{Main equations}

We will use the following notations: the effective action $W[\eta, \bar\eta, J_\mu]$ which is generator of all weakly connected Green functions,  is connected
with the generator of all Green functions $Z[\eta, \bar\eta, J_\mu]$ (partition function) as follows:
$$
Z[\eta, \bar\eta, J_\mu]=\exp\left(iW[\eta, \bar\eta, J_\mu]\right).
$$
Consequently, the vacuum expectations (in the presence of external sources) for quantum fields are
$$
\psi_i=\frac{\delta W}{\delta \bar\eta^i},\qquad  \bar\psi_i=-\frac{\delta W}{\delta \eta^i}\qquad\mbox{and}\quad A^\mu_i=\frac{\delta W}{\delta J_\mu^i}.
$$

Applying to the following classical action for QED (here all the functions are classical fields, not vacuum expectation):
\begin{equation}\label{clasqed}
    S=\frac{i}{2}\overline{\psi}\hat\partial\psi-\frac{i}{2}\partial_\mu\overline{\psi}\gamma^\mu\psi-m\overline{\psi}\psi-e\overline{\psi}\widehat{A}\psi+\frac12 A^\mu D^{-1}_{\mu\nu}A^\nu,\qquad D^{-1}_{\mu\nu}=g_{\mu\nu}\partial^2-\left(1-\frac{1}{\alpha}\right)\partial_\mu\partial_\nu,
\end{equation}
the DeWitt operator \cite{dw}
\begin{equation}\label{dewitt}
    \Lambda=1-\frac{i\hbar}{2}\frac{\delta^2W}{\delta J_\mu^i\delta J_\nu^j}\frac{\delta^2}{\delta A^\mu_i\delta A_j^\nu}+i\hbar\frac{\delta^2W}{\delta J_\mu^i\delta \eta^j}\frac{\delta^2}{\delta A^\mu_i\delta \bar\psi_j}-i\hbar\frac{\delta^2W}{\delta J_\mu^i\delta \bar\eta^j}\frac{\delta^2}{\delta A^\mu_i\delta \psi_j}+
    i\hbar\frac{\delta^2W}{\delta \eta^i\delta \bar\eta^j}\frac{\delta^2}{\delta \psi_j\delta \overline{\psi}_i}+ \cdots
\end{equation}
in accordance with DeWitt equations
$$
\Lambda\frac{\delta S}{\delta A_\mu^i}=-J^\mu_i,\qquad
\Lambda\frac{\delta S}{\delta \psi^i}=\bar\eta_i,\qquad
\Lambda\frac{\delta S}{\delta \bar\psi^i}=-\eta_i
$$
it may be derived \cite{fm}
 the following set of functional equations for the QED effective action $W$ (here all the fields are vacuum expectations of quantum fields):
\begin{equation}\label{qed1}
   \begin{array}{l}
\displaystyle{-e\bar{\psi}\gamma^\mu\psi+D^{-1\mu\nu}A_\nu+ie\hbar \,\mathrm{Tr}\left(\gamma^\mu\frac{\delta^2W}{\delta\overline{\eta}\delta\eta}\right)=-J^\mu;}\\  \\
\displaystyle{\left[\bar\psi(x)\left( i{\overleftarrow{\widehat\partial}}+e\widehat{A}+m\right)\right]_\alpha+ie\hbar\left(\frac{\delta^2W}{\delta J_\mu^i\delta \eta^i}\gamma_\mu\right)_\alpha=\overline{\eta}_\alpha;}\\ \\
\displaystyle{\left[\left(i\widehat{\partial}-e\widehat{A}-m\right)\psi\right]_\alpha+ie\hbar\left(\gamma_\mu\frac{\delta^2W}{\delta
J_\mu^i\delta \bar\eta^i}\right)_\alpha=-\eta_\alpha.}
\end{array}
\end{equation}
This set of equations coincides with Schwinger equations \cite{bs}, of course.
Taking into account following relations (all derivatives over the
grassmann variables are understand as left ones)
\begin{equation}\label{dwdadpsi}
    \frac{\delta^2 W}{\delta J_\mu^i\delta\eta^j}=-\frac{\delta \bar\psi^j}{\delta J_\mu^i}=\frac{\delta A^\mu_i}{\delta \eta^j};\qquad  \frac{\delta^2 W}{\delta J_\mu^i\delta\bar\eta^j}=\frac{\delta \psi^j}{\delta J_\mu^i}=\frac{\delta A^\mu_i}{\delta \bar\eta^j};\qquad  \frac{\delta^2 W}{\delta \overline{\eta}^\alpha\delta\eta^\beta}=-\frac{\delta \overline{\psi}^\beta}{\delta\overline{\eta}^\alpha}=-\frac{\delta \psi^\alpha}{\delta \eta^\beta}
\end{equation}
Eqs.(\ref{qed1}) may be bringed to the  more convenient for further
consideration  form:
\begin{equation}\label{qedeq}\begin{array}{l}
\displaystyle{ie\hbar\, \mathrm{Tr}\left(\frac{\delta \bar\psi^i}{\delta\bar\eta^i}\gamma_\mu\right)=ie\hbar\, \mathrm{Tr}\left(\frac{\delta \psi^i}{\delta\eta^i}\gamma_\mu\right)= J_\mu^i-e\bar{\psi}\gamma_\mu\psi+D^{-1}_{\mu\nu}A^\nu ;  }\\ \\
\displaystyle{       -ie\hbar\frac{\delta \bar\psi^i}{\delta J_\mu^i} \gamma_\mu=ie\hbar \frac{\delta \widehat{A}}{\delta \eta}=\bar\eta_\alpha^i- \left[\bar\psi(x)\left( i{\overleftarrow{\widehat\partial}}+e\widehat{A}+m\right)\right]_\alpha                                                    ;}\\ \\
\displaystyle{     ie\hbar\gamma_\mu\frac{\delta \psi^i}{\delta
J_\mu^i} =   ie\hbar  \frac{\delta \widehat{A}}{\delta\bar \eta}=
-\eta^i_\alpha-
\left[\left(i\widehat{\partial}-e\widehat{A}-m\right)\psi\right]_\alpha
.}
\end{array}\end{equation}
Usual way to solve these equations is perturbative expansion over
small coupling constant $e.$ I.e.,
 at first step we put coupling constant $e=0$, thereby turning these equations into  equations for free particles which can be solved easily.
 Further, interactions between free particles are taken
into account iteratively over small parameter $e$ (really, $e^2/4\pi$) getting
perturbative power series:
\begin{equation}\label{pertW}
    W=W_0+eW_1+e^2W_2+\cdots.
\end{equation}
This is usual way. But there is one circumstance must be taken into
account. It is obvious that in front of each derivative term there
is coupling constant $e$  (and $\hbar$), this means if we put $e=0$
then our (variational) differential  equations transform to
algebraic ones. In the following section we show that in this case
a full solution to equations of such type (with  small parameter)  should
contains not only regular    but singular  (in $e$) part too.
In the Sect.\ref{singsec} we will adopt this method to QED vacuum expectations.

\section{The method of solution}
The formulation of problem  under consideration may be explained in
the following simple example \cite{vb}: find solution to equation
\begin{equation}\label{mudadt}
    \mu\frac{dx(t)}{dt}=a(t)x(t)+b(t),\qquad x(0)=x_0, \qquad 0\leq t <\infty
\end{equation}
in the form of perturbative expansion over small $\mu.$ It is very
easy to find exact solution to this equation obeying given boundary
condition:
\begin{equation}\label{xott}
    x(t)=x_0 \exp\left(\frac{1}{\mu}\int\limits_0^ta(s) ds\right)+\frac{1}{\mu}\int\limits_0^tb(s)\exp\left(-\frac{1}{\mu}\int\limits_t^sa(z)dz\right)ds.
\end{equation}
It is obvious that  $\mu=0$  is essential singular point for
solution to (\ref{mudadt}) and, consequently, regular perturbative
expansion for small $\mu$  can not exist. The reason for such
situation can be seen from Eq.(\ref{mudadt})  itself: if we put
$\mu=0$ in this equation then it fails to be differential equation
and becomes to be algebraic one
\begin{equation}\label{munol}
    a(t)\tilde{x}(t)+b(t)=0.
\end{equation}
But solution to this (algebraic) equation $\tilde{x}(t)=-b(t)/a(t)$
in general can not obey given boundary condition: $a(0)/b(0)\neq
x_0.$ It happens loss of boundary condition.  This means that
solution to Eq.(\ref{munol}) can not be considered even as first
approximation to exact solution of Eq.(\ref{mudadt}). This
consideration underlies the reason for singularity at $\mu=0.$

From above mentioned  it follows that any expansion of a solution to
Eq.(\ref{mudadt}) around $\mu=0$ may be singular one only.
Derivation of a singular perturbation series according to \cite{vb}
consist of the following steps. First, take the second term in
(\ref{xott}) and integrate it by parts to get the following series:
\begin{multline*}
\frac{1}{\mu}\int\limits_0^tb(s)\exp\left(-\frac{1}{\mu}\int\limits_t^sa(z)dz\right)ds=-\left[\frac{b(t)}{a(t)}+\mu\left(\frac{b(t)}{a(t)}\right)^\prime\frac{1}{a(t)}+\cdots\right]+\\
+\left[\frac{b(0)}{a(0)}+\mu\left(\frac{b(0)}{a(0)}\right)^\prime\frac{1}{a(0)}+\cdots\right]\exp\left(\frac{1}{\mu}\int\limits_0^ta(s)
ds\right).
\end{multline*}
As a result the following series is obtained:
\begin{multline*}
\displaystyle{x(t)=-\left[\frac{b(t)}{a(t)}+\mu\left(\frac{b(t)}{a(t)}\right)^\prime\frac{1}{a(t)}+\cdots\right]+}\\
\displaystyle{+\left[x_0+\frac{b(0)}{a(0)}+\mu\left(\frac{b(0)}{a(0)}\right)^\prime\frac{1}{a(0)}+\cdots\right]\exp\left(\frac{1}{\mu}\int\limits_0^ta(s)
ds\right).}
\end{multline*}
Let's to make substitution $t=\mu \tau,\,\,s=\mu\zeta$ in the
integrand of the exponent. Then
$$
\frac{1}{\mu}\int\limits_0^ta(s) ds=\int\limits_0^\tau
a(\mu\zeta)d\zeta=a(0)\tau+\mu\frac{a^\prime(0)}{2}\tau^2+\mu^2\frac{a^{\prime\prime}(0)}{6}\tau^3+\cdots
$$
or,
$$
\exp\left(\frac{1}{\mu}\int\limits_0^ta(s)
ds\right)=e^{a(0)\tau}\left[1+\mu\frac{a^\prime(0)}{2}\tau^2+\mu^2\frac{a^{\prime\prime}(0)}{6}\tau^3+\mu^2\frac{\tau^4}{4}a^{\prime
2}(0)+\cdots\right]
$$
So it has been derived the following series over $\mu:$
$$
x(t, \mu)=\tilde{x}(t)+\Pi x(\tau),
$$
where
\begin{equation}\label{regpart}
    \tilde{x}(t)=\tilde{x}_0(t)+\mu \tilde{x}_1(t)+\cdots=-\frac{b(t)}{a(t)}-\mu\left(\frac{b(t)}{a(t)}\right)^\prime\frac{1}{a(t)}+\cdots
\end{equation}
is a \textit{regular} part of the solution and

  \begin{equation}\label{singpart}
    \Pi x(\tau)=\Pi_0x(\tau)+\mu\Pi_1x(\tau)+\mu^2\Pi_2x(\tau)+\cdots
\end{equation}
  is a \textit{singular} one with
  \begin{equation}\label{sngprt}
    \Pi_0x(\tau)=\left(x_0+\frac{b(0)}{a(0)}\right)e^{a(0)\tau};\qquad \Pi_1x(\tau)=\left[\left(x_0+\frac{b(0)}{a(0)}\right)a^\prime(0)\frac{\tau^2}{2}+\left(\frac{b(0)}{a(0)}\right)^\prime\frac{1}{a(0)}\right]e^{a(0)\tau}
  \end{equation}
  etc.
The terms $\Pi_k x(\tau)$  are called \textit{boundary layer} terms.
It is easy to see that
\begin{equation}\label{boundcond}
    \left(\tilde{x}_0(t)+\Pi_0x(\tau)\right)_{t=0}=x_0;\qquad \left(\tilde{x}_k(t)+\Pi_kx(\tau)\right)_{t=0}=0,\quad k\geq1.
\end{equation}
Now we can present the algorithm of singular perturbative solution
in the following form. Given an equation with boundary condition
(\ref{mudadt}). Then the solution should be divided into two parts
as follows: $x(t)=\tilde{x}(t)+\Pi x(\tau)$ and Eq.(\ref{mudadt})
can be presented in a form:
\begin{equation}\label{alg1}
\mu\frac{d\tilde{x}}{dt}+\frac{d\Pi x(\tau)}{d\tau}=a(t)
\tilde{x}(t)+a(\mu \tau)\Pi x(\tau)+b(t), \qquad \tau=t/\mu.
\end{equation}
Further one should to expand each term in both sides of this
equation in series over $\mu$. Equating coefficients in front of the
same powers of $\mu$, separately for terms depending on $t$ and
terms depending on $\tau$, one obtains equations for determination
of terms of the expansions (\ref{regpart})  and (\ref{singpart}). It
is easy to check out that in this way the series (\ref{regpart})
and (\ref{sngprt}) will be obtained. And it is not so hard to see,
that this algorithm is equivalent to consider Eq.(\ref{alg1}) as
divided in to two parts - first part includes terms depending on
$t$, and second part includes terms depending on $\tau$. Solutions
of these equations connect each other through boundary conditions
(\ref{boundcond}).

\section{The regular part of the effective action}
Let's to introduce new scaled variables $\rho=\eta/e, \,
\bar\rho=\bar\eta/e$  and $j_\mu=J_\mu/e.$ And then present each
field in Eq.(\ref{qedeq}) in the form, divided into regular and
singular parts:
$$
\psi=\psi^R(J, \eta, \bar\eta; e)+\Pi\psi(ej, e\rho,\,e\bar\rho;
e),\qquad \overline{\psi}=\overline{\psi}^R(J, \eta, \bar\eta;
e)+\Pi\overline{\psi}(ej, e\rho,\,e\bar\rho; e),
$$
$$
 A_\mu=A_\mu^R(J, \eta, \bar\eta; e)+\Pi A_\mu(ej, e\rho,\,e\bar\rho; e).
$$
Further acting in accordance with above mentioned (in the end of
preceding section) method one should to divide Eqs.(\ref{qedeq})
into part depending on  $J, \eta, \bar\eta$  and part, depending on
$j,  \rho,  \bar\rho.$ Let's for simplification of equations denote
the sources as follows: $J, \eta, \bar\eta \Leftrightarrow s$  and
scaled sources as follows: $j, \rho, \bar\rho \Leftrightarrow
\sigma.$ These settings allows one to write down equations in more
shorter form because now it is possible to denote: $\psi^R(J, \eta,
\bar\eta; e)=\psi^R(s; e)$ for regular part and $\Pi\psi(J, \eta,
\bar\eta; e)=\Pi\psi(e\sigma; e)$  for singular part of the field
$\psi$. The same notations  will be used for other fields too. Then
equations for regular parts will be (for sake of simplicity in the
belove equations for regular parts we will omit the superscript
$R$):
\begin{equation}\label{regequat}\begin{array}{l}
\displaystyle{ie\hbar\, \mathrm{Tr}\left(\frac{\delta \overline\psi^{\,i}(s; e)}{\delta\overline\eta^i}\gamma_\mu\right)=ie\hbar\, \mathrm{Tr}\left(\frac{\delta \psi^i(s; e)}{\delta\eta^i}\gamma_\mu\right)= J_\mu^i-e\overline{\psi}(s; e)\gamma_\mu\psi(s; e)+D^{-1}_{\mu\nu}A^\nu(s; e) ;  }\\ \\
\displaystyle{       -ie\hbar\frac{\delta \overline\psi^i(s; e)}{\delta J_\mu^i} \gamma_\mu=ie\hbar \frac{\delta \widehat{A}(s; e)}{\delta \eta}=\overline\eta_\alpha^i- \left[\overline\psi(s; e)\left( i{\overleftarrow{\widehat\partial}}+e\widehat{A}(s; e)+m\right)\right]_\alpha                                                    ;}\\ \\
\displaystyle{     ie\hbar\gamma_\mu\frac{\delta \psi^i(s;
e)}{\delta J_\mu^i} =   ie\hbar  \frac{\delta \widehat{A}(s;
e)}{\delta\overline \eta}=   -\eta^i_\alpha-
\left[\left(i\widehat{\partial}-e\widehat{A}(s; e)-m\right)\psi(s;
e)\right]_\alpha                                                 .}
\end{array}\end{equation}
Solutions to these equations will be searched in the regular
perturbation form:
\begin{equation}\label{regpert}\begin{array}{l}
    \psi(s, e)=\psi_0(s)+e\psi_1(s)+e^2\psi_2(s)+\cdots; \quad \overline{\psi}(s, e)=\overline{\psi}_0(s)+e\overline{\psi}_1(s)+e^2\psi_2(s)+\cdots;\\ \\
    A_\mu(s, e)=A_{0\mu}(s)+eA_{1\mu}(s)+e^2A_{2\mu}(s)+\cdots.
\end{array}\end{equation}
It is very simple to find these series by iterations. Equations for
zeroth order terms:
\begin{equation}\label{zero}
J_\mu^i+D^{-1}_{\mu\nu}A^\nu_0(s) =0;\quad \overline\eta-
\overline\psi_0(s)\left(
i{\overleftarrow{\widehat\partial}}+m\right)      =0;\quad
 -\eta^i_\alpha- \left(i\widehat{\partial}-m\right)\psi_0(s)=0      .
\end{equation}
Their solutions:
\begin{equation}\label{regsol0}
    A_{0\mu}(s)=-D_{\mu\nu}J^\nu;\qquad \overline\psi_0(s)=\bar\eta\frac{1}{i{\overleftarrow{\widehat\partial}}+m};\qquad \psi_0(s)=-\frac{1}{i\widehat{\partial}-m}\eta.
\end{equation}
Regular part of zeroth order effective action:
\begin{equation}\label{w0}
    W_0=-\bar\eta\frac{1}{i\widehat{\partial}-m}\eta-\frac12 J^\mu D_{\mu\nu}J^\nu.
\end{equation}

Equations for first order terms:
$$
\overline\psi_1(s)\left(
i{\overleftarrow{\widehat\partial}}+m\right)+\overline\psi_0(s)\widehat{A}_0(s)=0;\quad
\left(i\widehat{\partial}-m\right)\psi_1(s)-\widehat{A}_0(s)\psi_0(s)=0;
$$
$$
i\hbar
\mathrm{Tr}\left(\frac{1}{i\widehat{\partial}-m}\gamma_\mu\right)+\overline{\psi}_0\gamma_\mu\psi_0(s)=D^{-1}_{\mu\nu}A^\nu_1(s).
$$
Solutions to them:
\begin{equation}\label{regsol1}\begin{array}{l}
 \displaystyle{   \psi_1(s)=-\frac{1}{i\widehat{\partial}-m}\gamma_\mu\frac{1}{i\widehat{\partial}-m}\eta D^{\mu\nu}J_\nu;\quad \overline{\psi}_1=-\bar\eta(i{\overleftarrow{\widehat\partial}}+m)^{-1}\widehat{A}_0(i{\overleftarrow{\widehat\partial}}+m)^{-1};}\\ \\
\displaystyle{
A_{1\mu}(s)=-\bar\eta(i{\overleftarrow{\widehat\partial}}+m)^{-1}\gamma_\mu\frac{1}{i\widehat{\partial}-m}\eta+i\hbar
D_{\mu\nu}\mathrm{Tr}\left(\frac{1}{i\widehat{\partial}-m}\gamma^\nu\right).}
\end{array}\end{equation}
Regular part of first order effective action:
\begin{equation}\label{w1}
    W_1=-\bar \eta \frac{1}{i\widehat{\partial}-m}\gamma_\mu\frac{1}{i\widehat{\partial}-m}\eta D^{\mu\nu}J_\nu+i\hbar J^\mu D_{\mu\nu}\mathrm{Tr}\left(\frac{1}{i\widehat{\partial}-m}\gamma^\nu\right).
\end{equation}
Eq.(\ref{w0})  reproduces free propagators for electron-positron
and  photon fields. Eq.(\ref{w1}) reproduces first order Feynman
diagrams, including one-loop tadpole diagram. Acting this way one can
to reproduce the Feynman diagrams of all order in coupling constant (electric charge) $e$. This is why we have
called this series as regular perturbation ones. But the existence of the coupling constant (electric charge)
 $e$ in front of terms with derivatives in Eqs.(\ref{qedeq})
 set one thinking about possible singularity at $e=0.$

\section{Singular parts of vacuum expectations in QED\label{singsec}}
Equations for singular (boundary layer) parts are more complicated:
\begin{multline}\label{eqtnssing1}
\displaystyle{-i\hbar \mathrm{Tr}\left(\frac{\delta\Pi\psi(e\sigma;e)}{\delta\rho}\gamma_\mu\right)=-i\hbar \mathrm{Tr}\left(\frac{\delta\Pi\overline{\psi}(e\sigma;e)}{\delta\overline{\rho}}\gamma_\mu\right)=   e\Pi\overline{\psi}(e\sigma;e)\gamma_\mu\psi(e\sigma;e)  +}\\
\displaystyle{+e\overline{\psi}(e\sigma;e)\gamma_\mu\Pi\psi(e\sigma;e)+e\Pi\overline{\psi}(e\sigma;e)\gamma_\mu\Pi\psi(e\sigma;e)
-D^{-1}_{\mu\nu}\Pi A^\nu(e\sigma;e);            }
\end{multline}
\begin{multline}\label{eqtnssing2}
\displaystyle{                -i\hbar\frac{\delta \Pi\overline{\psi}(e\sigma;e)}{\delta j_\mu}\gamma_\mu= i\hbar\frac{\delta \Pi\widehat{A}(e\sigma;e)}{\delta \rho}=-\Pi\overline{\psi}(e\sigma;e)\left( i{\overleftarrow{\widehat\partial}}+m\right)-e\Pi\overline{\psi}(e\sigma;e)\widehat{A}(e\sigma;e)  -                         }\\
\displaystyle{
-e\overline{\psi}(e\sigma;e)\Pi\widehat{A}(e\sigma;e)-e\Pi\overline{\psi}(e\sigma;e)\Pi\widehat{A}(e\sigma;e);                                          }
\end{multline}
\begin{multline}\label{eqtnssing3}
\displaystyle{                i\hbar\gamma_\mu\frac{\delta \Pi\psi(e\sigma;e)}{\delta j_\mu}= i\hbar\frac{\delta \Pi\widehat{A}(e\sigma;e)}{\delta \overline{\rho}}=-\left( i\widehat\partial-m\right)\Pi\psi(e\sigma;e)+e\widehat{A}(e\sigma;e)\Pi\psi(e\sigma;e)  +                         }\\
\displaystyle{+e\Pi\widehat{A}(e\sigma;e)\psi(e\sigma;e)+e\Pi\widehat{A}(e\sigma;e)\Pi\psi(e\sigma;e);                                          }
\end{multline}
Recall that fields $\psi(e\sigma;e),  \,\overline{\psi}(e\sigma;e)$
and $A_\mu(e\sigma;e)$  are regular parts of corresponding fields,
but they are functions not of $s=(J_\mu, \eta, \bar\eta)$  but of
$e\sigma=(ej_\mu, e\rho, e\bar\rho.)$

At first step we should to extract zeroth order (in $e$) equations from above
mentioned ones:
\begin{equation}\label{zeroth1}
\displaystyle{i\hbar
\mathrm{Tr}\left(\frac{\delta\Pi_0\psi(\sigma)}{\delta\rho}\gamma_\mu\right)=i\hbar
\mathrm{Tr}\left(\frac{\delta\Pi_0\overline{\psi}(\sigma)}{\delta\overline{\rho}}\gamma_\mu\right)=
D^{-1}_{\mu\nu}\Pi_0 A^\nu(\sigma);            }
\end{equation}
\begin{equation}\label{zeroth2}
             -i\hbar\frac{\delta \Pi_0\overline{\psi}(\sigma)}{\delta j_\mu}\gamma_\mu= i\hbar\frac{\delta \Pi_0\widehat{A}(\sigma)}{\delta \rho}=-\Pi_0\overline{\psi}(\sigma)\left( i{\overleftarrow{\widehat\partial}}+m\right);
\end{equation}
\begin{equation}\label{zeroth3}
  i\hbar\gamma_\mu\frac{\delta \Pi_0\psi(\sigma)}{\delta j_\mu}= i\hbar\frac{\delta \Pi_0\widehat{A}(\sigma)}{\delta \overline{\rho}}=-\left( i\widehat\partial-m\right)\Pi_0\psi(\sigma).
\end{equation}
It is easy to find a general form of solutions to equations for
$\Pi_0\psi(\sigma)$   and $\Pi_0\overline{\psi}(\sigma)$:
\begin{equation}\label{pipsipipsi}\begin{array}{c}
   \Pi_0\psi(\sigma)=\exp\left[\frac{i}{4\hbar}\widehat{j}(i\widehat{\partial}-m)\right]f+c\psi_D;\\ \\
   \Pi_0\overline{\psi}(\sigma)=\overline{f}\exp\left[-\frac{i}{4\hbar}\left( i{\overleftarrow{\widehat\partial}}+m\right)\widehat{j}\right]+c\bar{\psi}_D,
\end{array}\end{equation}
where $\psi_D$ and $\bar{\psi}_D$ are solutions to free Dirac equations, $c$ - an arbitrary constant, $f$ - some spinor field.

 Now it is the time to apply  equation for boundary condition (\ref{boundcond}). From Eq.(\ref{regsol0}) follows that regular parts of fields under consideration
 vanish  at $J_\mu=\eta=\bar{\eta}=0.$ Further, from Lorentz invariance it follows that
vacuum expectation for spinor field in external source-free case vanishes:
$$
\psi\Big|_{J_\mu=\eta=\bar{\eta}=0}=\overline{\psi}\Big|_{J_\mu=\eta=\bar{\eta}=0}=0.
$$
These conditions give us that $f\Big|_{J_\mu=\eta=\bar{\eta}=0}=-c\psi_D.$ In general let's present the function $f$ as follows:
\begin{equation}\label{ffunction}
    f=-c\psi_D+\sum\limits_{n=0}^{\infty}c_n (\bar{\rho}\rho)^{n+s}\rho,
\end{equation}
where $s$ should be found from corresponding indicial equation (after defining of corresponding differential equation for $f$).
After substitution of Eq.(\ref{ffunction}) into  (\ref{pipsipipsi}) we have following expression for singular part of the spinor field:
\begin{multline}\label{singpsi}
 \Pi_0\psi(\sigma)=c\left[1-\exp\left[\frac{i}{4\hbar}\widehat{j}(i\widehat{\partial}-m)\right]\right]\psi_D+\exp\left[\frac{i}{4\hbar}\widehat{j}(i\widehat{\partial}-m)\right]\sum\limits_{n=0}^{\infty}c_n (\bar{\rho}\rho)^{n+s}\rho=\\
 =\exp\left[\frac{i}{4\hbar}\widehat{j}(i\widehat{\partial}-m)\right]\sum\limits_{n=0}^{\infty}c_n (\bar{\rho}\rho)^{n+s}\rho=\exp\left[\frac{i}{4e\hbar}\widehat{J}(i\widehat{\partial}-m)\right]\sum\limits_{n=0}^{\infty}c_n (\bar{\rho}\rho)^{n+s}\rho
\end{multline}
and conjugate expression for $\Pi_0\overline{\psi}(\sigma)$. These expressions has essential singularity at zero coupling limit $e\rightarrow 0.$
 Conclusion about essential singularity at zero coupling limit $e\rightarrow 0$
 can be referred to $\Pi_0 A^\nu$ too.

 So, any vacuum expectation in QED has essential singularity at zero coupling limit $e\rightarrow 0.$

\section{Conclusion}
The singularity at $e=0$ is very interesting - its existence means that in general we can't "switch off" electromagnetic interaction  and go to "free electron".
 It is the time to remember Dyson's proof \cite{ds} that perturbation series in QED is asymptotic one.
 Our consideration shows that QED effective action can't be an analytic function in the neighborhood of $e=0$  , consequently, any series in this region can't be convergent one.
    In the light of this singularity the notion of "free electron" should be revised - because it is impossible to "switch off" the electromagnetic interaction the existence of free, noninteracting electrically charged particle is questionable. But this point is very hard one and more accurate studies required to be conclusively established.

\end{document}